# Reply to Comment on "Symbolic Calculation of Two-Center Overlap Integrals Over Slater-Type Orbitals"

Sedat Gümüş and Telhat Özdoğan*
*Department of Physics, Faculty of Education, Ondokuz Mayis University, 05189 Amasya, Turkey*

The comments of Guseinov are critically analyzed. Contrary to his comments, it is pointed out that our formula for two-center overlap integrals over Slater type orbitals have been derived independently, not derived from the earlier works of Guseinov by changing the summation indices. Therefore, our algorithm is original, is not affected from possible instability problems and can be used in large scale calculations without loss of significant figures. Meanwhile, it should be stressed that his comment on the transformation of our formula into his formula proves the correctness of our algorithm and therefore can be regarded as a nice sound of science.

The comment on "Symbolic Calculation of Two-Center Overlap Integrals Over Slater-Type Orbitals" by I. I. Guseinov is analyzed. In his comment, he claims that the formulae we presented in Ref. 1 were not original and could be derived from his previous works[2-4] by changing the summation indices.

As is well known, the problem in the evaluation of multicenter integrals is to give an accurate and speed algorithm, which is not affected from possible instabilities in the cases of nearly equal or equal orbital exponents and higher or lower internuclear distances. For avoiding such instability problems, several methods have been used in literature. One of the methods is to express Legendre functions in such a way that the algorithm does not suffer from the possible instability problems. Therefore, Rodrigues' formula for Legendre functions was first given by D. M. Silver and K. Ruedenberg[5] and applied to the evaluation of multicenter integrals many years ago before the works of Guseinov.[2-4]

The first aim of our recent paper[1] was to give an accurate and speed algorithm for the evaluation of two-center overlap integrals over STOs using the Rodrigues' formula for Legendre functions (this formula given by Eq. (5) and cited with Ref. 17 in our paper[1]). It can be seen from the tables in our paper[1] that the use of Rodrigues' formula for Legendre functions enables us to compute two-center overlap integrals efficiently, accurately and also without loss of figures for arbitrary quantum numbers, orbital exponents and internuclear distances. Therefore, we point out that our algorithm is not affected from the possible instability problems and can also be used in large scale calculations. For the instability problems occurring in the evaluation of multicenter integrals, it is advised to read the excellent paper of Barnett[6] on the *digital erosion* occurring in some methods [i.e. Ref. 7].

The second aim of our recent paper[1] was to give some symbolic tables for two-center overlap integrals over STOs since symbolic results can be highlighted to the relative benefits of all different methods. The importance of the symbolic calculation of multicenter integrals can be seen in review paper of Barnett.[8]

Contrary to Guseinov comments, we think that any physical or mathematical quantity obtained in two different ways can be transformed into each other. In this respect, the comments of Guseinov on the transformation of our formulae into his formulae prove that our algorithm is correct and can be regarded as a sound and nice contribution to the science. As a result, the formulae in our paper[1] are original and not derived from the earlier works of Guseinov by changing the summation indices.

* Corresponding author. E-mail: telhatoz@omu.edu.tr